\documentclass[twocolumn,longbibliography,superscriptaddress]{revtex4-1}
\usepackage[latin1]{inputenc}
\usepackage[english]{babel}
\usepackage{lineno,hyperref}
\usepackage{bm,graphicx,graphics,amsmath,amssymb,epsfig,color}
\usepackage{dcolumn}

\newcommand{\ve}{\varepsilon}
\newcommand{\mA}{\mathcal{A}}

\newcommand{\mC}{\mathcal{C}}
\newcommand{\mD}{\mathcal{D}}
\newcommand{\mO}{\mathcal{O}}
\newcommand{\mH}{\mathcal{H}}

\newcommand{\mZ}{\mathcal{Z}}

\newcommand{\be}{\begin{equation}}
\newcommand{\ee}{\end{equation}}
\newcommand{\bea}{\begin{eqnarray}}
\newcommand{\eea}{\end{eqnarray}}
\newcommand{\bse}{\begin{subequations}}
\newcommand{\ese}{\end{subequations}}
\newcommand{\comment}[1]{}
\newcommand{\gcol}[1]{{\color{black} #1}}
\newcommand{\ggcol}[1]{{\color{black} #1}}
\newcommand{\rlcol}[1]{{\color{black} #1}}

\begin{document}
\title{Condensation transition and ensemble inequivalence\\  in the Discrete Nonlinear Schr\"odinger Equation}

\author{Giacomo Gradenigo}
\affiliation{Gran Sasso Science Institute, Viale F. Crispi 7, 67100 L'Aquila, Italy}
\email{giacomo.gradenigo@gssi.it}

\author{Stefano Iubini}
\affiliation{Consiglio Nazionale delle Ricerche, Istituto dei Sistemi Complessi, via Madonna del Piano 10, 
  I-50019 Sesto Fiorentino, Italy}
\affiliation{Istituto Nazionale di Fisica Nucleare, Sezione di Firenze, via G. Sansone 1 I-50019,
  Sesto Fiorentino, Italy}

\author{Roberto Livi} \affiliation{Dipartimento di Fisica e Astronomia
  and CSDC, Universit\`a di Firenze, via G. Sansone 1 I-50019, Sesto
  Fiorentino, Italy} \affiliation{Istituto Nazionale di Fisica
  Nucleare, Sezione di Firenze, via G. Sansone 1 I-50019, Sesto
  Fiorentino, Italy} \affiliation{Consiglio Nazionale delle Ricerche,
  Istituto dei Sistemi Complessi, via Madonna del Piano 10, I-50019
  Sesto Fiorentino, Italy}

\author{Satya N. Majumdar}
\affiliation{LPTMS, CNRS, Universit\'e Paris-Sud, Universit\'e Paris-Saclay, 91405 Orsay, France}






\begin{abstract}

  The thermodynamics of the discrete nonlinear Schr\"odinger equation
  in the vicinity of infinite temperature is explicitly solved in the
  microcanonical ensemble by means of large-deviation techniques. A
  first-order phase transition between a thermalized phase and a
  condensed (localized) one occurs at the infinite-temperature
  line. Inequivalence between statistical ensembles characterizes the
  condensed phase, where the grand-canonical representation does not
  apply. The control over finite size corrections of the
  microcanonical partition function allows to design an experimental
  test of delocalized negative-temperature states in lattices of cold
  atoms.

\end{abstract}

\maketitle

\section{Introduction}
The Discrete Nonlinear Schr\"odinger equation (DNLSE)~\cite{KEVR} is a
very useful phenomenological model for various physical phenomena,
e.g.  light propagating in arrays of optical waveguides \cite{ESMBA}
and Bose-Einstein condensates in optical lattices \cite{TS,FLOP}.  In
fact, it can be viewed as a lattice version of the Gross-Pitaevskii
equation (nonlinear Schr\"odinger equation), i.e. the semiclassical
model for the bosonic condensate wave-function~\cite{PS16}.
In this paper, we solve the statistical mechanics of this model close
to infinite temperature, where it exhibits a peculiar first-order
phase transition from a thermalized phase to a localized/condensed
one.  So far, the condensation phenomenon in the DNLSE has been
investigated in depth only numerically~\cite{FLOP,LFO,ICOPP,IFLOP}.
It occurs when a macroscopic fraction of energy is localized on
one/few lattice sites and it can be traced back to the existence of
two conserved quantities~\cite{SEM14,Chatt}.
As we are going to make clear in what follows, a grand-canonical
description of this condensation transition is not allowed, because of
the inequivalence of statistical ensembles.  The overall method for
estimating the microcanonical partition function in the vicinity of
the phase transition relies on well established large-deviation
techniques.  Moreover, our results apply in any spatial dimension.

Before entering \rlcol{into} technical details, we want to point out that our
contribution can be cast in the more general frame of a deeper
understanding of the mechanisms responsible for wave-function
localization in systems of interacting quantum particles. This is one
among the most exciting open problems nowadays, both in condensed
matter and in statistical physics~\cite{NH15,AP17,A18}. As opposed to
the well known phenomenon of Many-Body Localization, taking place when
interactions are weakly non-linear, we show here that localization in
the DNLSE is essentially due to strong nonlinear effects. 
Notice that this phenomenon is
not due to the proximity to a \emph{standard} integrable system,
e.g.~\cite{BCP13,RMS15}, as usually invoked for many non-linear
models, nor it can be attributed to the effect of disorder, as in most
models of glasses. Here we are rather facing and characterizing quite
a unique and unprecedented scenario: lack of \rlcol{energy equipartition}, driven by
genuine strong nonlinearities.

\section{Model and state of the art}
The \ggcol{one-dimensional} parameter-less DNLSE Hamiltonian  is defined in terms of a
scalar complex field $z_j$, taking values on a lattice of $N$ sites
with periodic boundary conditions:
\be \mH =
\sum_{j=1}^N(z_j^*z_{j+1}+z_j z^*_{j+1}) + \sum_{j=1}^N |z_j|^4 \,,
\label{eq:Hamiltonian}
\ee \ggcol{where quadratic terms represent standard hopping energies
  between nearest-neighbour sites, while non-linear terms represent an
  on-site potential coming from the repulsive contact
  energy in the Gross-Pitaevskii equation.}
The corresponding  equations of motion are
\be 
i \dot{z_j} = - \frac{\partial \mH}{\partial z_j^*} =
-(z_{j+1}+z_{j-1})- 2 |z_j|^2 z_j \,.
\label{eq:eq-motion}
\ee

One can easily check that this dynamics conserves $\mH$ and the {\sl total mass}
$\mA = \sum_{j=1}^N |z_j|^2$.  We denote by $E$ and $A$ the real values taken
by the observables $\mH$ and $\mA$, respectively. 

\ggcol{The thermodynamic behaviour of the DNLSE depends on the values
  taken by these two parameters or by their averages, depending on
  whether the system is isolated (microcanonical conditions) or
  interacting with an external reservoir (grand-canonical conditions).
  Accordingly, it is expected that the two conserved quantities can be
  mapped to a couple of values of temperature $T$ and chemical
  potential $\mu$.}

\begin{figure}
  \includegraphics[width=0.9\columnwidth]{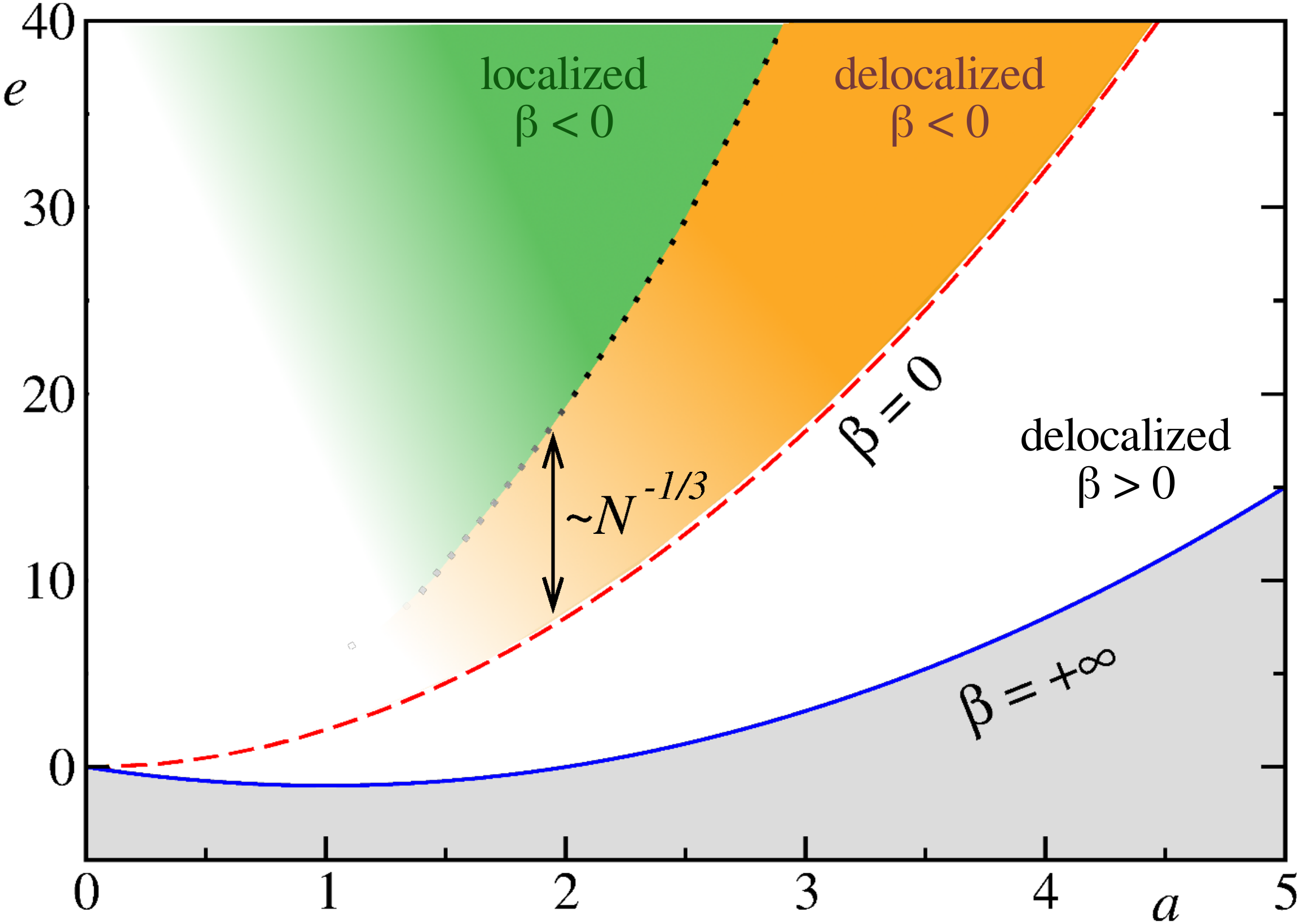}
  \caption{Equilibrium \ggcol{microcanonical} phase diagram of the
    DNLSE in the $(a, e)$-plane.  Blue solid and red dashed lines
    correspond to the ground state and the infinite temperature
    isothermal ($\beta=0$), respectively.  Localized (green) and
    \ggcol{delocalized negative-temperature} (orange) regions are
    drawn for a chain of $N=100$ lattice sites. Black dots identify
    the line of critical energy densities, see text.  }
\label{fig1}
\end{figure}

The main thermodynamic features of this model (e.g., see \cite{RCKG00}) are summarized in the phase
diagram shown in Fig.~\ref{fig1}, where $e=E/N$ and $a=A/N$ are the energy and mass densities,
respectively. 

\ggcol{In~\cite{RCKG00} it was shown within the grand-canonical
  ensemble that } the condition $e=2a^2$ identifies the
infinite--temperature line ($\beta = 1/T =0$ \ggcol{and
  $\mu^{-1}=0$}), while the ground state line ($\beta=+\infty$) is
identified by the relation $e=a^2-2a$ \ggcol{ and $\mu=2(a-1)$}. Below
this line no physical state is allowed, while a standard thermal phase
extends in between these two lines: dynamics (\ref{eq:eq-motion})
makes any initial state eventually thermalize to energy equipartition.
Above the $\beta = 0$ line numerical studies (e.g., see
\cite{FLOP,ICOPP}) have shown that the hamiltonian dynamics of the
DNLSE exhibits \ggcol{localized breather excitations}.  Such
excitations can survive over extremely long times (exponentially long
in their mass ~\cite{ICOPP}) even in a chain made of a few tens of
sites~\cite{FLOP}. The coupling of the localized breathers among
themselves, through the background energy fluctuations, is very weak
and their coalescence into a single localized giant breather, even if
possible (in fact, breathers are known to merge when they collide
\cite{FLOP}), is practically unobservable. This process can be
dramatically accelerated by replacing the hamiltonian dynamics with a
stochastic evolution rule, conserving both $\mH$ and $\mA$. As shown
in \cite{IPP14,IPP17} this stochastic dynamics yields full
localization as a product of a coarsening process.

In Ref.~\cite{RCKG00} it was also conjectured that the region above
the $\beta = 0$ line should correspond to a negative-temperature
phase. A numerical evidence of this conjecture was presented
in~\cite{JR,LFO,IFLOP}, although no conclusive theoretical argument
for the existence of this phase has been presented so far.  On the
other hand, several authors, on the basis of general thermodynamic
arguments, have argued that this negative-temperature phase should not be a
genuine thermodynamic equilibrium one (e.g., see
\cite{RN,R1,R2,R3,R4,BM18,CEF}). In this paper we provide an answer to
all of these open questions, by computing explicitly the
microcanonical partition function of the DNLSE close to the $\beta =
0$ line for large $N$, using large-deviation techniques similar to
those employed in \cite{SEM14,SEM14b,GB17,GM19}. \gcol{In particular,
  the study of the microcanonical entropy here presented closely
  follows the large deviations study of run-and-tumble particles
  in~\cite{GM19}, though the context of the latter was entirely
  different.}

\section{Statistical Mechanics of the DNLSE}
We start by pointing out that close to the infinite temperature line
the \rlcol{scalar complex fields in their polar representation, $z_j=
  \rho_j e^{i\phi_j}$, have random phases,} \ggcol{ so that in the
  limit of large values of $N$ the bilinear hopping term in
  (\ref{eq:Hamiltonian}) is of order $\sqrt{N}$, being the sum of
  random sign numbers, while the nonlinear contribution to energy is
  of order $N$, being the sum of positive numbers. That is why at high
  temperature hopping terms yield a subleading contribution to energy
  and the $\beta=0$ line is determined by the relation $e=2a^2$, even
  according to the exact calcultion of~\cite{RCKG00} where hopping
  terms are retained. The latter result really indicates that at high
  temperatures the total energy reduces to the contribution of the
  local quartic potential. In the following calculations, we will thus
  always neglect hopping terms. While a more thorough discussion on
  this random phase approximantion can be found in~\cite{IFLOP,GILM},
  here it is worth \rlcol{anticipating} one of our results, namely
  that in the limit $N\rightarrow\infty$ the microcanonical
  temperature diverges for all energies $e>2a^2$. This furnishes an
  \emph{``\`a posteriori''} argument to accept the random-phase
  approximantion as a reasonable one for the whole region $e>2a^2$ of
  the phase diagram. We stress that the results hereafter reported
  hold for any value of $d$, since any relation with the spatial
  dimension $d$ enters this problem through a $d$-dimensional hopping
  term $\sum_{k=1}^d (z_{\vec{j}}^* z_{\vec{j}+\vec{e}_k} +z_{\vec{j}}
  z_{\vec{j}+\vec{e}_k}^*)$, where $\vec{j}\in \mathbb{Z}^d$ and
  $\{\vec{e}_k\}$ with $k=1,\cdots,d$ is the unit vector basis for
  $\mathbb{Z}^d$.} To our knowledge this is the first theoretical
model exhibiting a \rlcol{localization transition, induced by
  nonlinearity,} in finite dimensions $d>1$.\\

Since $A$ and $E$ are the only two conserved quantities, the
microcanonical partition function \rlcol{ for $\beta\to 0$, where the hopping term can be neglected,} reads:
\be \Omega_N(A,E) = (2\pi)^N \int \mD \rho ~\delta(A -
\sum_{j=1}^N \rho_j^2)~\delta(E-\sum_{j=1}^N \rho_j^4),
\label{eq:microcanonical_2}
\ee
where we have used \ggcol{the short-hand notation $\int \mD\rho =
  \int_0^{\infty}\prod_{j=1}^N d\rho_j~\rho_j$ and we have integrated
  over the phases $\phi_j$}.  It is convenient to compute first the
Laplace transform $\tilde\Omega_N(\lambda,E)$:
\begin{align}
  \tilde\Omega_N(\lambda,E) = \int_0^\infty dA~e^{-\lambda A} \Omega_N(A,E)
\end{align}
By performing the change of variables $\rho_j^4=\ve_j$, this can be
rewritten as follows
\be \tilde\Omega_N(\lambda,E)=\left(\frac{\pi}{\lambda}\right)^N~\int\mD \ve
\,f_\lambda(\ve_j)~\delta\left( E -\sum_{j=1}^N \ve_j \right),
\label{eq:canonical-1}
\ee
with
\be 
\int \mD \ve = \int_0^{\infty} \prod_{j=1}^Nd\ve_j
\ee
and 
\be
f_\lambda(\ve) = \lambda/(2\sqrt{\ve})\exp\left( -\lambda\sqrt{\ve} \right)\,. 
\label{eq:fx}
\ee
The function $f_\lambda(\ve)$ can be recognized as a normalized
probability distribution with a {\em stretched-exponential} tail, with
average and variance
\be
\langle \ve \rangle = 2/\lambda^2\quad , \quad \sigma^2 = \langle \ve^2 \rangle - \langle \ve \rangle^2 = 20/\lambda^4\,.
\label{eq:lambda}
\ee
The integral on the r.h.s. of~(\ref{eq:canonical-1}) is formally
identical to the probability that the sum of $N$ independent
identically distributed random variables $\ve_j$, with individual
distribution $f_\lambda(\ve)$, takes the value $E$.  Large-deviation
theory predicts that, due to the stretched exponential tail of
$f_\lambda(\ve)$, such probability undergoes a condensation phenomenon
when $E$ overtakes the threshold value $E_{th} = N \langle \ve
\rangle$ \cite{EH2005,S08-leshouches,SEM14,SEM14b,GM19}
 : a finite fraction of the total energy localizes onto a single site
\ggcol{in the thermodynamic limit.}
Let us stress that the exponent of the slow
stretched-exponential decay of $f_\lambda(\ve)$ is uniquely determined
by the order of the non-linearity. For instance, if the non-linear
term had been $|z_j|^6$, one would have obtained $f_\lambda(\ve) \sim
\exp(-\ve^{1/3})$.

\begin{figure}
  \includegraphics[width=0.9\columnwidth]{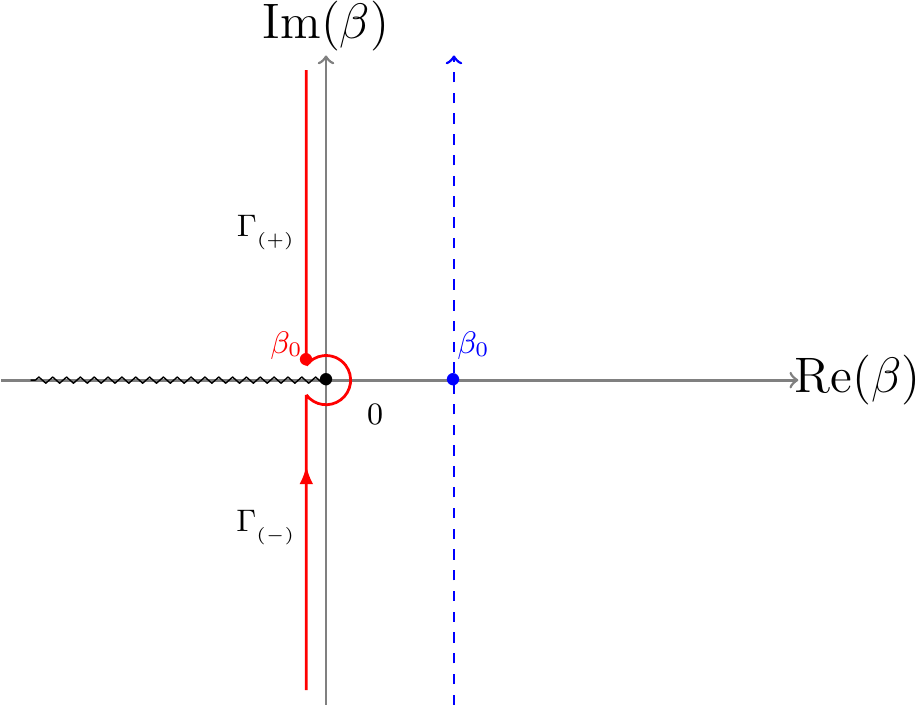}
  \caption{Analytic structure of $\mZ_N(\lambda,\beta)$ in the complex $\beta$ plane.
  The microcanonical partition function can be computed by the Bromwich contour
  represented by the blue line and by the red line for
  $E < E_{th}$ and $E > E_{th}$, respectively.}
\label{fig2}
\end{figure}

Following a standard procedure~\cite{SEM14,SEM14b,GB17,GM19}, one should compute the
Laplace transform of $\tilde\Omega_N(\lambda,E)$ with respect to $E$,
i.e.
\be
\mZ_N(\lambda,\beta)= \int_0^\infty dE~e^{-\beta
  E} \tilde\Omega_N(\lambda,E),
\ee
which can be rewitten in the following form
\be \mZ_N(\lambda,\beta)= (\pi/\lambda)^N \exp\left\lbrace N \log[z(\lambda,\beta)]
\right\rbrace,
\ee
where 
\be z(\lambda,\beta) =
\frac{\lambda}{\sqrt{\beta}}~e^{\lambda^2/(4\beta)}~\text{Erfc}\left(\frac{\lambda}{2\sqrt{\beta}}\right)
\, ,
\label{eq:canonical-2}
\ee
and $\text{Erfc}(x)=\int_x^\infty~e^{-t^2}~dt$ is the complementary
error function. For any fixed  $\lambda >0$, the expression in~(\ref{eq:canonical-2}) can be taken as the definition of a
function in the complex $\beta$ plane. In particular 
$\text{Erfc}[\lambda/(2\sqrt{\beta})]$, and accordingly $z(\lambda,\beta)$, 
 has a branch-cut on the
negative real semiaxis (see Fig.~\ref{fig2}).
Inverting formally this Laplace transform with respect to $\beta$, we obtain
$\tilde\Omega_N(\lambda,E)$ as 
\be
\tilde\Omega_N(\lambda,E) = \int_{\beta_0-i\infty}^{\beta_0+i\infty} \frac{d\beta}{2\pi i}~e^{\beta E}\, \mZ_N(\lambda,\beta)
\label{eq:Zeta-gamma}
\ee
For large $N$ the integral on the right hand side (r.h.s.) can be estimated by a saddle-point 
approximation at the real value $\beta_0$, 
which is the solution of the saddle-point equation
\be
E = - \frac{\partial \ln \mZ_N(\lambda,\beta)}{\partial \beta} \,.
\label{eq:saddle-point-z}
\ee
Taking into account the analytic properties of $\mZ_N(\lambda,\beta)$
one finds that a real positive solution $\beta_0$ exists only if $E <
E_{th}$ and the integral in~(\ref{eq:Zeta-gamma}) can be performed
over the vertical contour (dashed blue line in Fig.\ref{fig2})
intersecting the positive real axis in $\beta_0$. In particular, this
guarantees also the equivalence between the micro-canonical and the
grand-canonical statistical ensembles.  Conversely, for $E > E_{th}$ a
real solution of (\ref{eq:saddle-point-z}) does not exist, because the
integration contour cannot intersect the branch-cut along the real
negative semiaxis in the complex $\beta$-plane. Accordingly, there is
no equivalence between statistical ensembles.  This notwithstanding,
the microcanonical partition function can still be computed, because
the integral in~(\ref{eq:Zeta-gamma}) can be explicitly performed
by following a suitably deformed contour (continuous red line in
Fig.\ref{fig2}).  In practice, this demands the application of a large
deviation technique, which amounts to estimate
$\tilde\Omega_N(\lambda,E)$ at different orders of $E-E_{th}$ as a
power of $N$: details can be found in \cite{GILM}.  Here we just
summarize the main results. At the energy scale $E-E_{th}\sim
N^{\frac{1}{2}}$ (Gaussian regime), $\tilde\Omega_N(\lambda,E)$ can be
estimated by expanding $\mZ_N(\lambda,\beta)$ up to order $\beta \sim
N^{-\frac{1}{2}}$.  As an almost straightforward consequence of the
central limit theorem one obtains [see also~(\ref{eq:lambda})]
\be \tilde\Omega_N(\lambda,E)
= \frac{1}{\sigma\sqrt{2\pi N}}\exp\left[ - \frac{(E -
    E_{\text{th}})^2}{2\sigma^2 N} \right]\,.
\label{eq:ZE-Gaussian}
\ee
In the regime of extreme large deviations the energy scale is
$E-E_{th}\sim N$ and $\mZ_N(\lambda,\beta)$ has to be expanded up to
order $\beta \sim N^{-1}$, yielding
\be
\tilde\Omega_N(\lambda,E) \sim  \exp\left[ - \sqrt{E - E_{\text{th}}}\right]\,
\label{eq:ZE-large}
\ee
From previous studies on formally identical
models~\cite{SEM14,SEM14b,GM19}, we know that at the scale
$E-E_{th}\sim N^{1/2}$ there is no localized phase, while it is
present when $E-E_{th}\sim N$. The localization transition is expected
to take place at the energy scale $E-E_{th}\sim N^{\frac{2}{3}}$,
where the analytic and the non-analytic contributions to the expansion
of $z(\lambda,\beta)$ close to $\beta=0$ are of the same order~\cite{GILM}. In
practice, this amounts to expand at the scale $\beta \sim
N^{-\frac{1}{3}}$, thus yielding
\be
\tilde\Omega_N(\lambda,E) = \left(\frac{\pi}{\lambda}\right)^N \left[\frac{1}{\sigma\sqrt{2\pi N}} \exp{\left(- N^{1/3} \frac{\zeta^2}{2\sigma^2}\right)} + \mC(\lambda,\zeta)\right]\,
\label{eq:Z-splitting}
\ee
with 
\be
\mC(\lambda,\zeta) = 
N \sqrt{\frac{2\pi}{\langle \ve \rangle}} \exp{[- N^{1/3} \chi(\zeta)]}
\quad , \quad
\zeta = \frac{E-E_{\text{th}}}{N^{2/3}} \,.
\label{eq:scaling-variable}
\ee In the first term on the r.h.s. of (\ref{eq:Z-splitting}) one
can easily recognize the gaussian contribution
in~(\ref{eq:ZE-Gaussian}). The explicit expression of the function
$\chi(\zeta)$ is quite involved (see Appendix B in \cite{GILM}). Here,
we just report its asymptotic behaviours
\be \chi(\zeta)
= \begin{cases} \frac{3}{2}\left(\frac{\sigma}{\langle \varepsilon
    \rangle}\right)^{2/3} \qquad\qquad\qquad\qquad \zeta\rightarrow
  \zeta_l \\ \\ \sqrt{\frac{2}{\langle \varepsilon \rangle}}
  \sqrt{\zeta}-\frac{\sigma^2}{4\langle\varepsilon\rangle}\frac{1}{\zeta}+\mO\left(\frac{1}{\zeta^{5/2}}\right),
  \quad \zeta\gg 1 \end{cases},
\label{eq:chi-asymptotics-intro}
\ee
where
\be
\zeta_l = \frac{3}{2}~\left(\frac{\sigma^4}{\langle \varepsilon \rangle}\right)^{1/3}
\label{eq:spinodal-zl}
\ee
is the spinodal point for the localized phase~\cite{GILM}.

The final step for recovering an explicit expression of the microcanonical  partition function amounts to
computing the inverse Laplace-transform with respect to $\lambda$:
\bea
&&\Omega_N(A, E) = \frac{e^{N\log(\pi)}}{2\pi i}\int_{\lambda_0-i\infty}^{\lambda_0+i\infty} 
d\lambda~e^{N [a\lambda-\log(\lambda)]} \times \nonumber  \\
 && \left[ e^{- N^{1/3} \chi(\zeta)} + e^{- N^{1/3}\zeta^2/(2\sigma^2)} \right]\,,
\label{eq:Laplace-inverse-mu-matching}
\eea 
The leading contribution to this integral for large $N$ is 
given by the first exponential in the r.h.s., yielding the  solution $\lambda_0 = 1/a$
of the saddle-point equation
\be
\frac{\partial}{\partial \lambda} \left[ a \lambda - \ln \lambda \right] = 0 \quad  \to \quad \lambda_0 = \frac{1}{a}\,,
\label{eq:saddle-point-a} 
\ee
We finally obtain the expression
\be
\Omega_N(A,\zeta)~\approx~e^{N\left[ 1 + \log(\pi a) \right]}~\left[ e^{- N^{1/3} \chi(\zeta)} + e^{- N^{1/3}\zeta^2/(2\sigma^2)} \right],
\label{eq:micro-matching}
\ee
where $\sigma^2 = 20 a^4$ and $\langle \ve \rangle = 2 a^2$. The latter expression tells us that
$E_{th}/N = \langle \ve \rangle = 2 a^2$, i.e. the line at infinite temperature in Fig.~\ref{fig1}, coincides with
the energy density  threshold, above which the theory of large deviations predicts the existence of the
condensed phase \cite{GM19}. 

\begin{figure}
  \includegraphics[width=0.9\columnwidth]{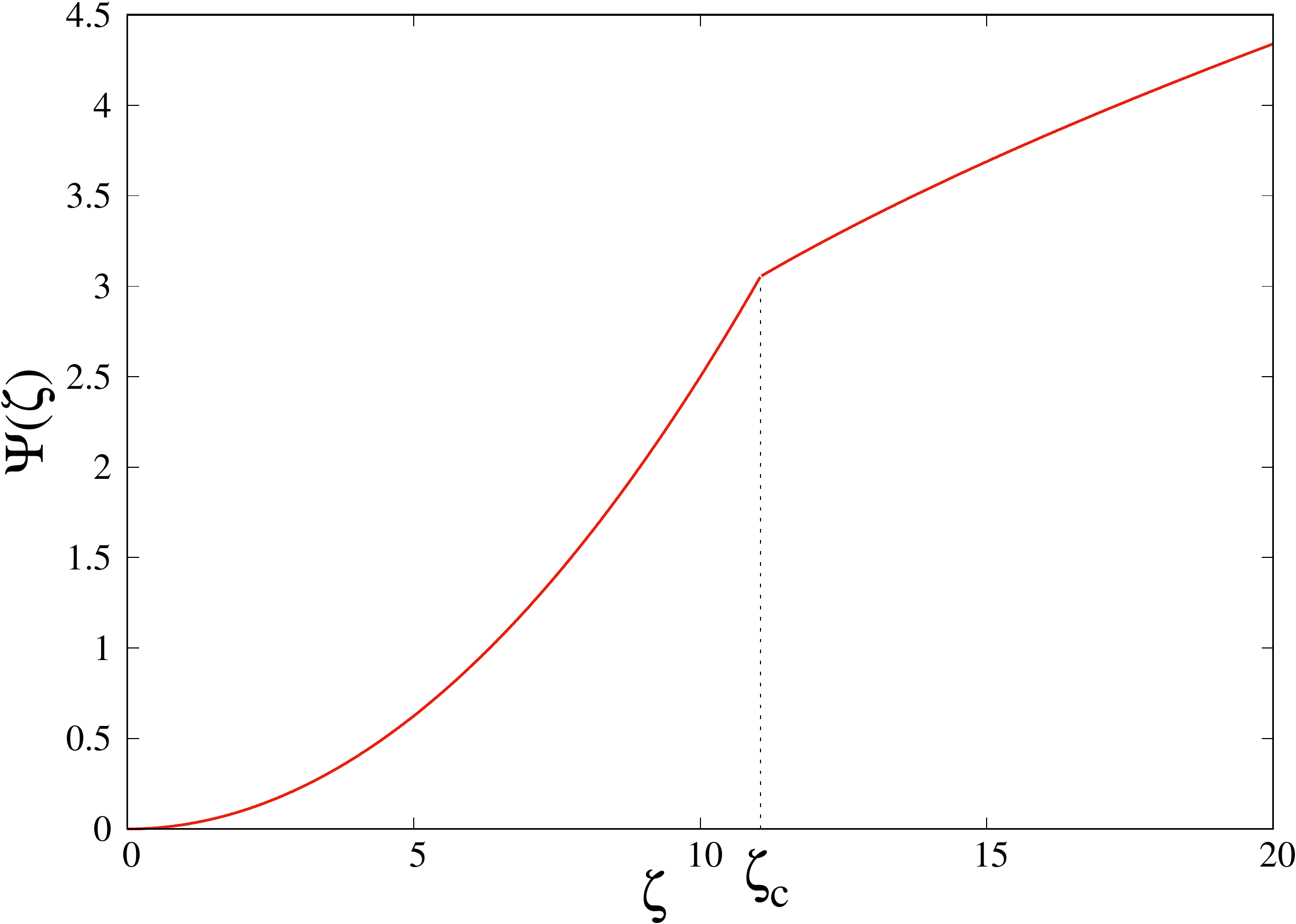}
  \caption{ Singular part $\Psi(\zeta)$ of the microcanonical entropy, see~(\ref{eq:rate-func-all}).
   Localization takes
    place at $\zeta_c\approx 11.05$, where $\Psi'(\zeta)$ is discontinuous.}
\label{fig3}
\end{figure}

In the limit of large values of $N$ we can thus write an estimate of the 
microcanonical entropy in the matching regime as follows
\be
S_N(a,\zeta) = N\left[ 1 + \log(\pi a) \right] - N^{1/3} \Psi(\zeta),
\label{eq:rate-func-all}
\ee
where 
\be 
\Psi(\zeta) = \text{inf}_{\zeta}\Big\{\chi(\zeta),  \zeta^2/(2\sigma^2)\Big\}\, .
\label{eq:rate-func}
\ee $S_N(a,\zeta)$ has an extensive part representing a background
energy--independent entropy, while the subleading contribution
contains information about the phase transition from a thermalized to
a condensed (localized) phase.  The critical value of $\zeta_c$
associated with the condensation transition is identified by the
condition \ggcol{that the two contributions in
  Eq.~(\ref{eq:rate-func}) are equal, i.e.}  $\chi(\zeta_c) =
\zeta_c^2/(2\sigma^2)$, yielding $\zeta_c =
2^{1/3}\,\zeta_l$~\cite{GILM}.  The function $\Psi(\zeta)$ is shown in
Fig.~\ref{fig3} and its derivative clearly exhibits a discontinuity at
$\zeta_c$, thus providing \ggcol{a first-order transition mechanism at
  large $N$. In order to make more explicit its dependence on energy
  for values above the threshold, $E>E_{th}$, the microcanonical
  entropy can be written as
\begin{align}
  S_N(A,E) = N [1+\log(\pi A/N)] - N^{1/3} \Psi(E).
  \label{eq:entropy-bis}
\end{align}
Clearly, any thermodynamic potential, computed taking derivatives of
$S_N(A,E)$ with respect to energy, does vanish for $E>E_{th}$ if we
consider only the term of leading order in $N$ in
Eq.~(\ref{eq:entropy-bis}). In such a regime any non trivial
dependence on $E$ comes from the subleading term $N^{1/3} \Psi(E)$. It
is for this reason that the discontinuity of $\Psi'(E)$ at $E=E_c$
dominates the physics for energies above the threshold $E_{th}$. In
particular we have that the contribution $N [1+\log(\pi A/N)]$ to
$S_N(A,E)$, in the regime $E>E_{th}$, yields just a constant prefactor
in front of the partition function, provided that one varies $E$
keeping $A$ fixed, a prefactor which simplifies in the expression of
any microcanonical expectation value. We will in fact see that the
participation ratio, which is defined as the average value of a certain
function over the microcanonical measure, has a discontinuity at
$E_c$.}

\ggcol{Note that for finite $N$ one can identify a critical value of
  the energy, $E_c = E_{th} + N^{2/3} \zeta_c$, which does not
  coincide with the threshold energy.}  The line of critical energy
densities $E_c(a)/N$ is the black dotted line shown in the phase
diagram of Fig.~\ref{fig1} for $N=100$. \ggcol{ Even for moderately
  large system sizes, the condensation transition line is
  substantially shifted to higher energy densities with respect to the
  line $E_{th}(a)/N$.  The energy density gap $E_c/N -E_{th}/N$
  vanishes as $N^{-1/3}$ in the thermodynamic limit.} Moreover, in the
region above $E_{th}$ the microcanonical inverse temperature
\begin{align}
  \frac{1}{T} = \frac{\partial S_N(A,E)}{\partial E} = - \frac{1}{N^{1/3}} \Psi'(E)
\end{align}
is negative both above and below $E_c$. The difference between these
two negative-temperature regions is characterized by the order parameter
of localization, the so-called participation ratio:
\be
Y_2 (E)= \left\langle \left(\sum_{j=1}^N \ve_j^2\right )\Bigg/
\left(\sum_{j=1}^N \ve_j \right)^2 \right\rangle_{A,E},
\label{eq:part-ratio}
\ee
where the angular brackets $\langle \cdot \rangle_{A,E}$ denote the
micro-canonical equilibrium average. When energy is democratically
distributed on all sites, i.e., $\ve_j \sim E/N$ for every $j$, one
has $Y_2 (E) \sim 1/N$. On the contrary, when a finite fraction of
energy is localized on a single site, i.e., $\ve_j \sim N$, then $Y_2
(E) \sim 1$. As shown in~\cite{GILM} $Y_2(E)$, can be computed
explicitly from~(\ref{eq:micro-matching}). For $E>E_c$ one obtains
that $Y_2 (E)$ is independent on $N$, thus identifying a localized
phase. For $E<E_c$ one finds $Y_2(E) \sim 1/N$, i.e., absence of
localization.

\ggcol{Altogether, for finite sizes we obtain the following physical
  scenario: the line $E_{th}$ separates the region at positive
  temperature, characterized by equivalence of ensembles and absence
  of localization from a region \rlcol{with} negative microcanonical temperature
  with ensembles inequivalence.  For $E_{th}<E<E_c$ we identify a
  peculiar delocalized phase at negative temperature, (see the orange
  region in Fig.~\ref{fig1}). For $E>E_c$ equilibrium states are
  localized and at negative temperature (green region in
  Fig.~\ref{fig1}).}  This result suggests that one could design an
experimental setup of cold-atoms in a finite optical lattice to
explore \ggcol{the delocalized region above $E_{th}$}, where true
negative-temperature states can be produced, while avoiding massive
localization effects, that could destabilize the atomic
condensate. Moreover, additional investigations should be devoted to a
precise characterization of 
the \ggcol{delocalized negative-temperature region}. In particular, it
would be interesting to compare \rlcol{it} with the weakly non-ergodic phase
described in~\cite{MKDF18}.

\section{Final remarks} 

In the thermodynamic limit, the above scenario dramatically
simplifies.  In fact $E_c/N$ and $E_{th}/N$ coincide, i.e. the
\ggcol{delocalized negative-temperature region} disappears and the
temperature turns out to be infinite above $E_{th}/N$.  Moreover, the
participation ratio vanishes continuously at $E_{th}/N$.  This
notwithstanding, the microcanonical partition function guarantees that
above $E_{th}/N$ genuine thermodynamic equilibrium states at infinite
temperature containing a unique localized breather do exist. Let us
also remark that the exponent of the finite-size scaling correction
$N^{1/3}$ of the microcanonical entropy~[see~(\ref{eq:rate-func-all})]
is uniquely, albeit non trivially, determined by the order of the
non-linear interactions of the DNLSE. Accordingly, we have made
explicit not only that localization in this model is essentially due
to nonlinearity, but also that the order of the nonlinearity, which is
a property of the microscopic interactions, directly affects a
macroscopic property of the system, i.e. the finite-size scaling
corrections of the entropy. Since experiments on bosonic condensates
in optical lattices are typically performed for relatively small
lattices~\cite{MO06}, the model discussed in this paper indicates the
possibility of novel and clear-cut predictions for laboratory tests.

\begin{acknowledgements}
We thank for interesting discussions M. Baiesi, S. Franz, L. Leuzzi,
G. Parisi, P. Politi, F. Ricci-Tersenghi, L. Salasnich,
A. Scardicchio, F. Seno and A. Vulpiani. G.G. acknowledges the
financial support of the Simons Foundation (Grant No.~454949, Giorgio
Parisi) and the hospitality of ``Sapienza'', University of Rome, for
the first stages of this work. S.I. acknowledges support from Progetto
di Ricerca Dipartimentale BIRD173122/17 of the University of Padova.
R.L. acknowledges partial support from project
MIUR-PRIN2017 \emph{Coarse-grained description for non-
equilibrium systems and transport phenomena} (CO-
NEST) n. 201798CZL
\end{acknowledgements}  


\end{document}